\documentclass[12pt, preprint]{aastex}

\usepackage{graphicx}
\shorttitle{microwave of flux rope}

\shortauthors{Wu et al.}

\usepackage{color}
\usepackage{amsmath}
\usepackage[normalem]{ulem}

\usepackage{lineno}

\usepackage{multirow}

\begin{document}

\title{Microwave imaging of a hot flux rope structure during the pre-impulsive stage of an eruptive M7.7 solar flare}

\author{Zhao Wu\altaffilmark{1}, Yao Chen\altaffilmark{1}, Guangli Huang\altaffilmark{2}, Hiroshi Nakajima\altaffilmark{3}, Hongqiang Song\altaffilmark{1}, Victor Melnikov
\altaffilmark{4}, Wei Liu\altaffilmark{5}, Gang
Li\altaffilmark{6}, Kalugodu Chandrashekhar\altaffilmark{1},
Fangran Jiao\altaffilmark{1}}

\altaffiltext{1}{Shandong Provincial Key Laboratory of Optical
Astronomy and Solar-Terrestrial Environment, and Institute of
Space Sciences, Shandong University, Weihai, Shandong 264209,
China; yaochen@sdu.edu.cn} \altaffiltext{2}{Purple Mountain
Observatory, Chinese Academy of Sciences (CAS), Nanjing, 210008,
China} \altaffiltext{3}{Nobeyama Radio Observatory, NAOJ, 462-2 Nobeyama,
Minamimaki, Minamisaku, Nagano 384-1305, Japan}
\altaffiltext{4}{Central Astronomical Observatory at
Pulkovo, Russian Academy of Sciences, Saint Petersburg 196140,
Russia} \altaffiltext{5}{W. W. Hansen Experimental Physics
Laboratory, Stanford University, Stanford, CA 94305, USA}
\altaffiltext{6}{Department of Space Science and CSPAR, University
of Alabama in Huntsville, Huntsville, AL 35899,
USA}

\begin{abstract}

Corona structures and processes during the pre-impulsive stage of
solar eruption are crucial to understanding the physics leading to
the subsequent explosive energy release. Here we present the first
microwave imaging study of a hot flux rope structure during the
pre-impulsive stage of an eruptive M7.7 solar flare, with the
Nobeyama Radioheliograph (NoRH) at 17~GHz. The flux rope is also
observed by the SDO/AIA in its hot passbands of 94 and 131~\AA{}.
In the microwave data, it is revealed as an overall arcade-like
structure consisting of several intensity enhancements bridged by
generally weak emissions, with brightness temperatures ($T_B$)
varying from $\sim$10,000~K to $\sim$20,000~K. Locations of
microwave intensity enhancements along the structure remain
relatively fixed at certain specific parts of the flux rope,
indicating that the distribution of emitting electrons is affected
by the large scale magnetic configuration of the twisted
flux rope. Wavelet analysis shows a pronounced 2-min
period of the microwave $T_B$ variation during the pre-impulsive
stage of interest. The period agrees well with that reported for
AIA sunward-contracting loops and upward ejective plasmoids
(suggested to be reconnection outflows). This suggests that both
periodicities are controlled by the same reconnection process that
takes place intermittently at a 2-min time scale. We infer that at
least a part of the emission is excited by non-thermal energetic
electrons via the gyro-synchrotron mechanism. The study
demonstrates the potential of microwave imaging in exploring the
flux rope magnetic geometry and relevant reconnection process
during the onset of solar eruption.
\end{abstract}


\section{Introduction}
Coronal flux ropes, i.e. magnetic structures consisting of
field lines twisted around each other, are believed to be one of
the major pre-eruptive agents of solar coronal mass ejections
(CMEs) with free magnetic energy to drive the eruption
\citep[e.g.][]{Low96,Chen96,Forbes00,Low01}. Their existence has
been revealed by latest observational studies using high-quality
imaging data of the Atmospheric Imaging Assembly
\citep[AIA/SDO:][]{Lemen12} on board the Solar Dynamics
Observatory \citep[SDO:][]{Pesnell12} via its high-temperature
passbands at 131~\AA\ ($\sim$11~MK) and 94~\AA\ ($\sim$7~MK)
\citep{Zhang12,Cheng13,Sun14,Song14,Song15}. It has been therefore
suggested that such a pre-eruption flux rope contains
plasmas as hot as $\sim$10~MK, possibly heated by an underlying
reconnection process. Yet, it remains elusive about how the
reconnection works to heat or even form the flux rope and whether
non-thermal energetic electrons are produced during the process.

An eruptive M7.7 flare took place on 19 July 2012, with valuable
observations that may contribute to resolve the above issue. The
event was well-observed by the SDO and Reuven Ramaty High Energy
Solar Spectroscopic Imager \citep[RHESSI:][]{Lin02} spacecraft, as
well as the ground-based microwave Nobeyama Radioheliograph
\citep[NoRH:][]{Nakajima94,Takano97}. A number of studies on the
event have been published, revealing many intriguing evolutionary
details of the flare
\citep[e.g.,][]{Krucker14,Liu13,Liu2013,Patsourakos13,Sun14,Mor14}.
It was established that the eruption was associated with a
pre-existing high-temperature flux rope structure according to
AIA/SDO. The flux rope was likely formed by a preceding C4.5 flare
that took place $\sim$7~hours earlier. The rope stayed and evolved in
the corona for several hours before its eventual eruption leading
to the M7.7 flare and a fast CME propagating at a speed $>$
1000~km s$^{-1}$ \citep{Patsourakos13}.

\citet{Liu2013} presented a comprehensive study on the event using
both SDO and RHESSI data. They found that the event belongs to the
well-known Masuda-type flare with a hard x-ray (HXR) source
high-lying in the corona above the flaring soft x-ray (SXR) arcade
\citep{Masuda94}. They examined the spacing difference between
X-ray sources at various energy bands, as well as the temperature
distribution with a forward-modelling differential
emission measure (DEM) analysis \citep{Asch13}, to infer sites of
plasma heating and electron acceleration induced by reconnection.
Along with other detailed measurements on contracting/shrinking
plasma loops and ejective plasmoids, they concluded that the
primary loci of heating and electron acceleration were in the
reconnection outflow regions, rather than the reconnection site
itself. In addition, they also found that the reconnection
outflows, as revealed by fast contracting loops and upward-flowing
plasmoids during the flare pre-impulsive stage, exhibited a
quasi-periodic occurrence rate at $\sim$2~mins. This last finding
is of close relevance to the study presented here.

Most previous studies of this event focused on the SDO and RHESSI
data, while the microwave data of the event recorded by NoRH at 17
and 34~GHz have not been paid enough attention. The coronal
microwave emission is contributed by both thermal bremsstrahlung
and non-thermal gyro-synchrotron mechanisms
\citep[e.g.,][]{Dulk85}. Emission given by the latter carries
valuable information about energetic electrons and magnetic field
strength and geometry. \citet{Liu13} and \citet{Mor14} did analyze
the NoRH microwave data to reveal the spatial and velocity
distributions of energetic electrons along loops of the major M7.7
flare, yet mainly during the impulsive stage. Here we analyze the
microwave data during the pre-impulsive stage focusing on the
relatively-weak yet significant microwave emission in the region
corresponding to the hot AIA flux rope structure. To our
knowledge, this is the first imaging study of a hot flux rope in
the microwave regime. The study reveals valuable information about
the flux rope structure and the generally weak reconnection
process during the flare pre-impulsive stage. These are
representative of major physical ingredients leading to the
eventual solar eruption. Section~2 introduces more relevant
details of the event, as already presented by other authors. The
third section shows our major results on the microwave emission
associated with the flux rope. The fourth section discusses the
nature of coronal microwave emission, which is followed by the
section of conclusions and discussion.

\section{Data analysis and overview of the M7.7 eruptive flare}\label{sec2}
The M7.7 flare occurred at 04:17~UT and peaked at 05:58~UT on July
19 2012. It was a limb event from NOAA AR 11520. The event can be
separated into pre-impulsive (04:17~UT to 05:16~UT) and impulsive
(05:16~UT to 05:43~UT) stages, according to the derivative of the
GOES SXR light curves and the HXR 25-50~keV profiles
\citep[c.f.,][]{Liu2013}. The emission from the northern flare
loop foot was partly occulted by the solar disk, allowing the
above-the-loop-top source of the HXR emission to be observed.
In Figure~1(a-c), we present the 131, 94, and 171~\AA\
images observed at $\sim$04:47~UT, during the pre-impulsive stage
to show the pre-existing flux rope structure. The rope is best
seen at 131~\AA\, and almost invisible at 171~\AA\, indicating it
contains high-temperature plasmas. An accompanying animation is
available online showing the evolution of the event. The yellow
dashed curves in Figure~1(a) delineate the inferred outer border
of the flux rope structure, within which some twisted structures
can be clearly seen. In the 131~\AA\ image, there exists a thin
vertical structure connecting the tip of the cusp-like flare loops
and the northern end of the flux rope (pointed by the white
arrow), which has been recognized as the high-temperature current
sheet along which reconnections took place \citep{Liu2013}.

In the pre-impulsive stage, the flux rope structure exhibited a
slow-rising motion (see the online animation). Reconnection
outflows in the form of downward contracting arcades and ejective
plasmoids have been observed along the current sheet structure
\citep{Liu13,Liu2013}. Of particular interest is the
quasi-periodic occurrence of these reconnection-induced features.
\citet{Liu2013} found that there were 29 downward contracting
arcades and 25 ejective plasmoids from 04:15 to 05:25~UT, i.e.,
the pre-impulsive phase (see their Figure~7). We repeated this
analysis, and show our result in Figure~2(a) with height-time maps
along slice S1 (see Figure~1(a)). The red and black dashed tracks
represent the upward and downward moving structures, respectively.
We found 24 contracting arcades from 04:20 to 05:15~UT and 14
eruptive plasmoids from 04:20 and 04:50~UT, yielding an average
occurrence rate of 2.3 and 2.1~mins. Both outflows have their
origin from the vertical current sheet structure that is indicated
by the central long-dashed line in Figure~2(a). The linear-fitting
speeds of these tracks are in a range of 10 - 60~km s$^{-1}$ for
the downward-moving structures and 50 - 400~km s$^{-1}$ for the
upward moving structures. These results are consistent with those
obtained by \citet{Liu2013}, indicating the intermittent nature of
the pre-impulsive reconnection. Here we will examine whether this
intermittency has imprints on the microwave emission that is
likely related to energetic electrons. Note that, according to
RHESSI the coronal HXR emission within the flux rope at this stage
is too weak to provide statistically-meaningful photon counts.

For this purpose, we use the NoRH data recorded during the
pre-impulsive stage of this event at both 17 and 34~GHz, with a 1s
cadence and beam size being $\sim$17~arcsecs for 17~GHz (shown in
Figure~3(a)) and 8.7~arcsecs for 34~GHz. According to the NoRH
analysis manual (http://solar.nro.nao.ac.jp/norh/doc/, (manual
version 3.3)), there are 3 different NoRH synthesis methods, named
as Hanaoka, Koshix, and Fujiki programs, with the Koshix one being
the most appropriate for analyzing diffuse sources such as the one
studied here. So we mainly use this program for our study. For
comparison, we also tried the Hanaoka program, and found our main
results are not affected.

In the pre-impulsive phase of the M7.7 flare, the
activities of the Sun were limited to AR 11520 (there was no other
bright active region) and the maximum brightness temperature
($T_B$) was roughly 30000 K. Therefore, the noise in the NoRH raw
images (before applying the CLEAN procedure) of the radio Sun is
essentially determined by the system noise, which is given by the
sum of the antenna temperature due to the quiet Sun and the
receiver noise temperature. According to \citet{Takano96}, the
system noise temperature for 1s data is $\sim$ 1000-1500 K for
17~GHz and $\sim$ 6000 K for 34~GHz. In the synthesis of 17 GHz
images, we set the CLEAN level to be 3000 K that is the default
value, and integrated 12 frames (12s) of NoRH raw images before
applying the CLEAN algorithm to further reduce the system noise.
For 34~GHz data we used the Hanaoka program (note the Koshix
program is not applicable to 34 GHz data) and found that in the
flux rope region of interest $T_B$ is considerably less than the
corresponding sensitivity level ($\sim$ 6000 K). Therefore, in the
following we only analyze the 17~GHz data of NoRH.

\section{ Microwave imaging of the pre-impulsive hot flux rope structure}\label{sec3}

Coronal microwave sources were present around and above the
cusp region following the earlier C4.5 flare, with enhanced activities
during the pre-impulsive stage of the M7.7 flare. In
Figure 1(d), we have over-plotted the flux rope regime depicted by
the dashed curves in Figure 1(a). It can be seen that some distinct
yet relatively weak microwave sources appear in the middle part of
the flux rope region. In Figure~3, we further present 4 snapshots
of the NoRH microwave images. See also the online animation
accompanying Figure~1 for the evolutionary process from 04:35~UT
to 04:55~UT. As mentioned, several similar bright structures
(referred to as microwave islands) with $T_B$ $\sim$10000 to over
20000~K are present within the AIA 131~\AA\ flux rope structure.
Between adjacent sources, relatively weaker emission ($T_B
\sim$10000 - 14000~K) exists. This forms an overall arcade-like
microwave structure threaded by the localized brightness maxima.
The number of these $T_B$ maxima varies from 3 to 5. Their $T_B$
changes rapidly over time, yet their locations seem to be
relatively fixed to specific parts of the large structure.

The arcade-like microwave structure is clearly separated from the
bright flaring loop source, as seen from the second, third and
forth columns of Figure~3, and the animation. The overall
microwave structure presents a gradually-rising motion. This
agrees with the very similar motion of the AIA flux rope
structure, as shown by the height-time maps along slice S2 (see
Figure~2(b)) (for AIA 131~\AA\ data) and 2(c) (for NoRH 17~GHz
data) from which we see that the trajectory of the
microwave source (indicated by a long-dashed line) lies in the
middle part of the flux rope structure. The average rising speed
of microwave structure is $\sim$28~km s$^{-1}$. This further
supports the idea that the microwave arcade-like structure corresponds to
the internal part of the hot flux rope observed in the AIA 131,
94~\AA\ passbands. Note that it is not possible to tell which
specific AIA structure in Figure~1 contains the microwave-emitting
energetic electrons, since both background and foreground
structures are projected together onto the plane of the sky.

The energetic (or thermal) electrons accounting for the microwave
emission are likely produced and filled into the flux rope by the
ongoing reconnection during the pre-impulsive stage. The
reconnection is revealed by the AIA-observed contracting arcades
and upward ejection of plasmoids from the current sheet structure,
according to previous studies \citep[e.g.,][]{Liu13,Liu2013}. In
order to find whether the intermittency associated with AIA
reconnection outflows (see Figure~2) have some imprint on the
microwave emission, we conduct wavelet analyses on the temporal
profiles of $T_B$ in the flux rope region.

To analyze the temporal variation of the microwave intensity, we
first average the $T_B$ values in Region R for the overall flux
rope region (defined in Figure~3), from 04:35~UT to 04:55~UT. The
normalized $T_B$ averages after subtracting the general variation
trend are presented as the red plot in Figure~4(a), and the
obtained wavelet and global power spectra in Figure~4(b), from
which we see a significant period of 2~mins which carries the
strongest spectral power. Almost the same 2-min period appears in
a similar wavelet analysis for the normalized average $T_B$ values
in Circle C1 which is selected to be centered around the middle
microwave island and moving outward along S2.

For comparison, we also conduct wavelet analysis with the data
representative of the quite-Sun microwave background in Circle C2.
See Figure~4(d) for the results. The power is much weaker and no
significant periodicity exists. This indicates that the above
2-min periodicity is physical and indeed associated with the
microwave flux rope structure.

Thus, we conclude that the microwave emission exhibits a
significant 2-min quasi-periodic oscillation. The period agrees
well with that reported for the AIA sunward-contracting loops and
upward ejective plasmoids (suggested to be reconnection outflows)
during the same period \citep[see Figure~2,][]{Liu2013},
indicating both quasi-periodicities have the same physical origins
given by the pre-impulsive reconnection along the thin vertical
current sheet.

\section{Discussion on the nature of the coronal microwave emission}\label{sec4}

The first issue we should clarify is about the physical
significance of the weak coronal microwave sources, i.e., whether
the signals are real or produced artificially by the NoRH image
synthesis procedure. As mentioned in Sec. 2, the NoRH system noise
of 1s data is $\sim$1000-1500 K, and we have integrated 12s of
NoRH raw images before applying the CLEAN algorithm to further
reduce the noise. With a CLEAN level set to be 3000 K, we suggest
that sources observed at 17 GHz with $T_B$ $>$3000 K are real.
The physical significance of these microwave sources are also
supported by their location and dynamics that are highly
correlated with the AIA structure.

Then we need to infer whether the microwave sources are
given by non-thermal energetic electrons via the gyro-synchrotron
emission or by thermal electrons through the bremsstrahlung
emission, or by a combination of both \citep{Dulk85}. This is in
general difficult since we cannot take direct measurements on
thermal and magnetic parameters of the corona sources. One
traditional way to resolve this issue is to examine the rough
microwave spectrum using the NoRH two-frequency data
\citep[e.g.,][]{Narukage14}, however, as mentioned the 34~GHz data
are below the NoRH sensitivity and not usable here. Nevertheless,
there still exists a qualitative approach to infer the nature of
the microwave emission.

It is well-known that the plasma temperature and emission
measure can be deduced with the DEM method. Given these
parameters, the thermal bremsstrahlung contribution can be
calculated with the following equation given by \citet{Dulk85} in
the optical thin regime,
\begin{equation}
    T_{B}=9.78 \times 10^{-3}\frac{\texttt{EM}}{\nu^2 T^{1/2}}\times(24.5+ln(T/\nu))
\end{equation}
where $T_B$~(K) is the brightness temperature, EM=$n^2L$ is the
total emission measure ($n$~(cm$^{-3}$) is the number density and
$L$~(cm) is the column depth), $T$~(K) is the plasma temperature,
and $\nu$~(Hz) is the frequency. For the event of study, two
groups of authors have done the DEM analysis using different
methods \citep{Liu2013,Sun14}. Here, we simply read the published
numbers of the DEM-weighted average $T$ and EM, without repeating
the DEM analysis. Comparing Figure 6 of \citet{Liu2013} and Figure
11 of \citet{Sun14}, we see that results of the two studies are
qualitatively similar, with the flux rope outlined by regions of
high $T$ and EM, yet both parameters are still significantly less
than those in the flare loops. Within the C1 region defined in
Figure 3(d), around 04:50 UT we read from \citet{Sun14} that $T$
is in a range of $\sim$ 6 - 8 MK and EM is in a range of $\sim$ 1
- 2.5 $\times 10^{28}$ cm$^{-5}$, and from \citet{Liu2013} that
$T$ varies from 3 - 6 MK and EM varies from 1 - 2 $\times 10^{28}$
cm$^{-5}$. With these numbers, we obtain that the thermal $T_B$ at
17 GHz varies in a range of 2000 - 6000 K. Assuming that there is not sufficient material cooler than these
temperature ranges to contribute significantly to the brightness
temperature, the NoRH brightness temperatures appear to require a
non-thermal contribution.

To further show that at least the oscillating part of the
microwave signals is non-thermal, we examine temporal variations
of light curves of different AIA passbands which are related to
thermal plasmas to see whether a similar 2-min oscillation exists
or not. If not, we may attribute that the microwave oscillation
reported above belongs to non-thermal emission of energetic
electrons. To do this, we plot in Figure 5 the light curves
observed with the AIA 131, 94, 335, 211, 193, and 171 \AA\
passbands from 04:35 to 04:55~UT in the flux rope region (upper
panel) and C1 region (lower panel). We see that in these EUV
channels, the light curves vary rather smoothly with most curves
declining gradually (due to the rise and expansion of the flux
rope) without any notable oscillations. This further supports that
at least the oscillating component of the microwave sources is
given by non-thermal energetic electrons, possibly through the
gryo-synchrotron emission mechanism \citep{Dulk85}.

\section{Conclusions and Discussion}\label{sec5}

Investigating pre-impulsive processes is of crucial importance to
understanding the physics leading to explosive energy release of
solar eruption. Here we present such a study using
microwave imaging data of a flux rope structure observed during
the pre-impulsive phase of the M7.7 flare on July 19 2012.
Previous studies on the same event already showed the flux rope is
of high-temperature and is connected to the flare reconnection
region. Based on these studies, we focus on the microwave data at
17~GHz obtained by NoRH to reveal the microwave characteristics of
the same structure. We find that the flux rope, while being hot in
general, contains microwave-emitting energetic electrons. The flux
rope, as viewed in the microwave, exhibits several local maxima of
emission intensities with relatively fixed locations. These maxima, ~i.e., microwave islands, are bridged by generally
weaker emission, forming an overall arcade-like structure.

We also show that the flux rope microwave sources
seem at least partly due to non-thermal energetic electrons,
possibly at energies of hundreds of keV
\citep[e.g.,][]{White11,Kundu01,Kundu04}. High localization of
these microwave sources indicates that the energetic electrons may
be affected or confined by the large scale magnetic structure
within the flux rope. The confinement is probably via the
well-known magnetic mirror effect if local magnetic enhancements
are present. These magnetic enhancements can be due to twisting
flux rope field lines and/or their interaction with nearby
magnetic structures. Thus, the microwave islands can be used to
infer sites of magnetic enhancements within the flux rope
structure, which are at present very difficult, if not impossible,
to measure directly.

It is also found that the microwave emission is characterized by a
2-min oscillation period. The period agrees well with the
occurrence rate of the AIA-observed contracting loops and ejective
plasmoids that are manifestations of reconnection outflows during
the pre-impulsive stage. This correlation indicates that the
microwave periodicity is associated with the reconnection that
takes place and releases energetic electrons intermittently every
$\sim$2-min interval. Note that oscillations with similar
periodicity during flares have been reported and attributed to
flare-released fast magneto-sonic wave trains
\citep[e.g.,][]{Ofman11,Liu12,Yang15}, or some external
modulations by disturbances associated with the sunspot
\citep{Sych09,Reznikova11} or the photosphere/chromosphere
oscillations \citep{Kis11}. These independent observational
findings, together with ours, point to an intermittent nature of
the flaring reconnections, not only during their impulsive stage
but also in the pre-impulsive stage.

\acknowledgements

This work was partly carried out on the Solar Data Analysis System
operated by the Astronomy Data Center in cooperation with the
Solar Observatory of the National Astronomical Observatory of
Japan. We are grateful to the SDO teams for making their data
available to us. Z. W thanks Qiang Hu and Xing Li for useful discussions. This work was supported by grants NSBRSF
2012CB825601, NNSFC 41274175 and 41331068. Gang Li's work at
UAHuntsivlle was supported by NSF grants ATM-0847719 and
AGS1135432.

\begin{figure}
\includegraphics[width=0.85\textwidth]{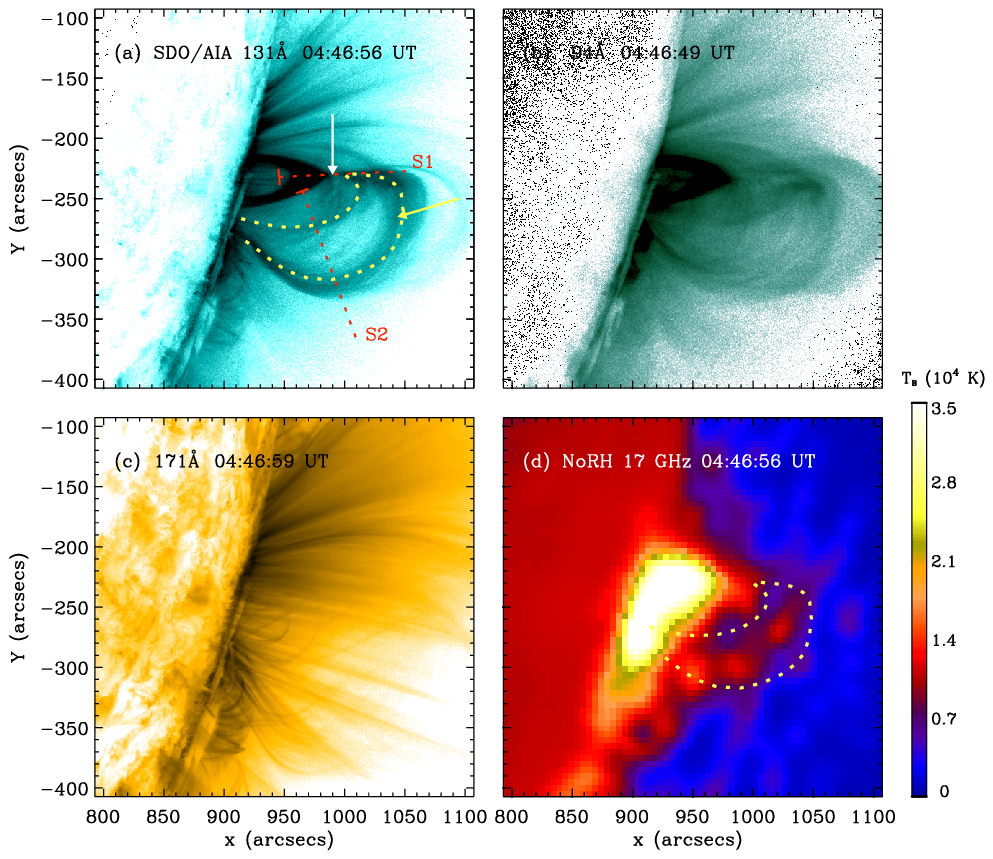}
\caption{AIA and NoRH images at $\sim$04:56 of the pre-impulsive
phase of the M7.7 flare on July 19 2012. Panel (a)-(c) present
images in passbands of 131, 94, and 171~\AA\ and panel (d)
presents the image of NoRH at~17 GHz. The yellow dashed curves in
(a) delineate the flux rope structure in 131~\AA\, which have been
superposed onto the microwave image in (d). The white arrow in (a)
points to the vertical thin current sheet structure. Slices S1 and
S2 are used to plot the height-time maps shown in Figure~2. The
short bars at the inner end of S1 and S2 indicate the starting
point of distance measurement along the slice. \protect\\(An
accompanying animation and a color version of this figure are
available in the online journal.)}\label{Fig1}
\end{figure}

\begin{figure}
\includegraphics[width=0.68\textwidth]{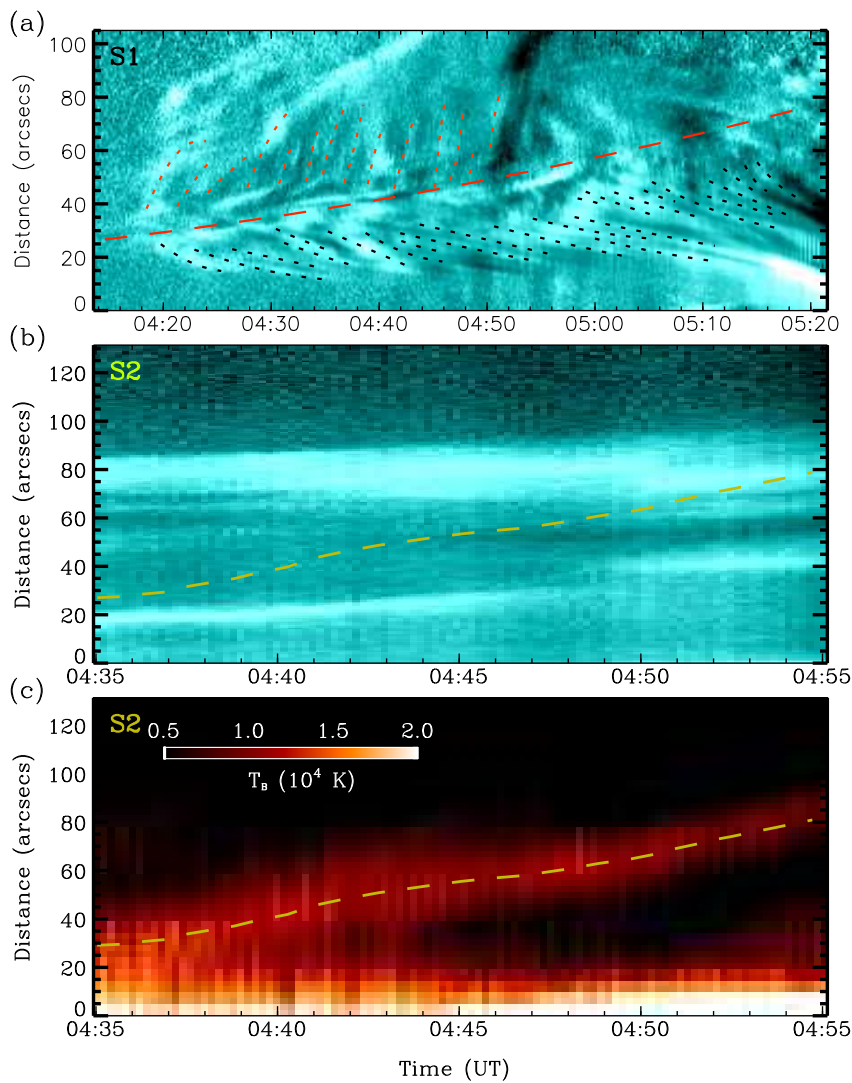}
\caption{Height-time maps along Slices S1 and S2. See Figure~1 for
slice locations. Panel (a) is for AIA 131~\AA\ to show the
reconnection outflows, observed from 04:14~UT to 05:22~UT.
Downward-contracting loops are indicated by black dashed lines,
upward moving plasmoids by red dashed lines. The middle red
long dashed line shows the reconnecting current sheet region.
Panels (b) and (c) are height-time maps along S2 for AIA 131~\AA\
and the brightness temperature data of NoRH at 17~GHz,
respectively, from 04:35~UT to 04:55~UT. The NoRH data are
synthesis images processed by the Koshix program with 12~s
integration of the raw data. Dashed lines in (b) and (c) indicate
the rising motion of the microwave structure.\protect\\(A color
version of this figure is available in the online
journal.)}\label{Fig2}
\end{figure}

\begin{figure}
\includegraphics[width=0.75\textwidth]{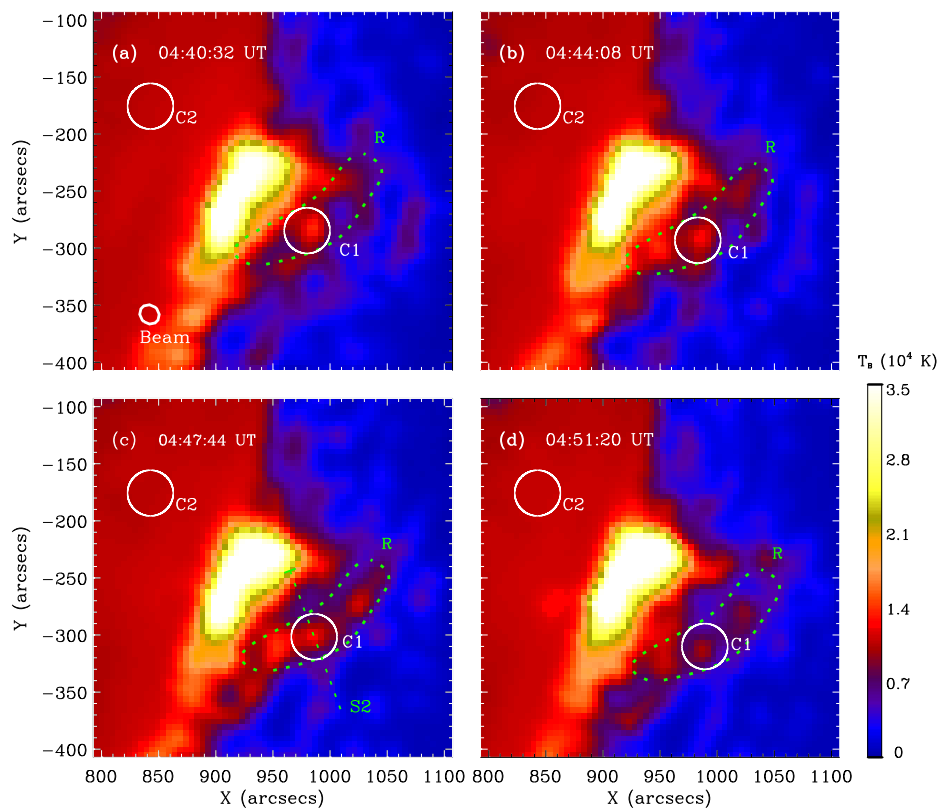}
\caption{Image sequence of the NoRH 17~GHz microwave data from
04:40~UT to 04:52~UT, showing the brightness temperature ($T_B$).
Region R indicates the overall microwave flux rope regime observed
at 17 GHz, and Circle~C1 (C2) denotes the region surrounding the
middle microwave intensity enhancement (the solar background) with
radius $\sim$20~arcsecs. Both region R and Circle~C1 move outward
along S2 with a projection speed of $\sim$28~km s$^{-1}$. The beam
size of NoRH 17~GHz is shown in panel~(a).\protect\\(A color
version is available in the online journal. The accompanying
animation has been presented with Figure~1.)} \label{Fig3}
\end{figure}

\begin{figure}
\includegraphics[width=0.65\textwidth]{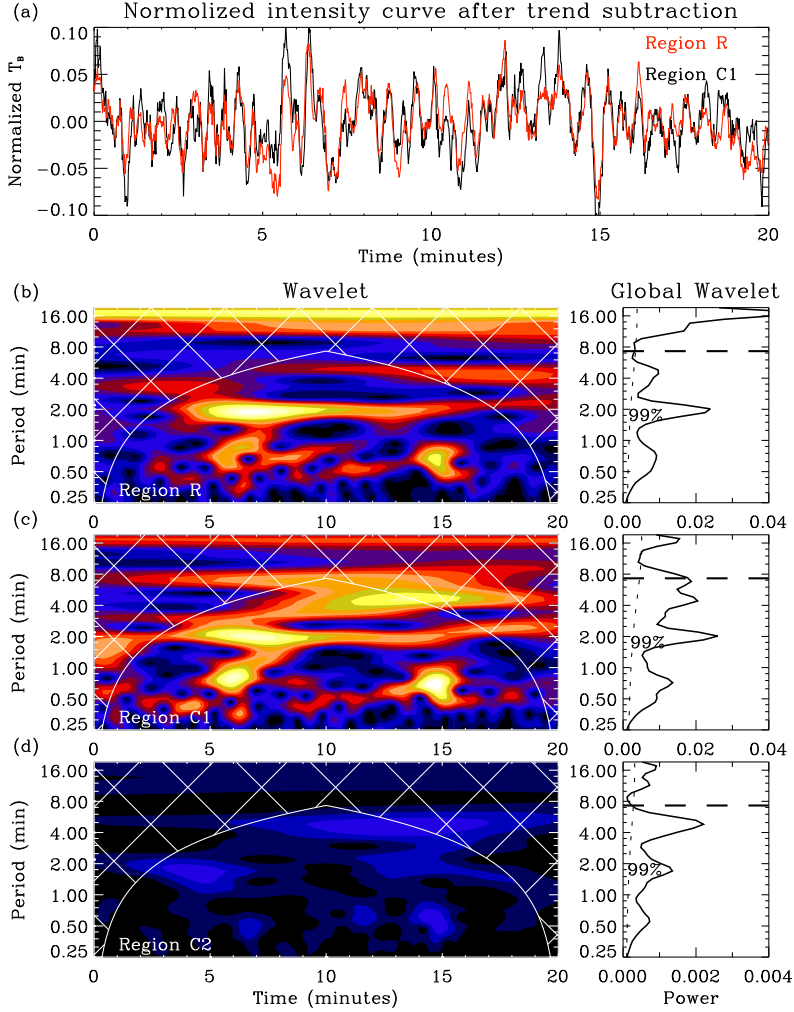}
\caption{Wavelet analysis of the NoRH 17~GHz $T_B$ temporal
variations. Panel (a) shows the averaged $T_B$ (normalized after
trend subtraction) in Region R and C2 from 04:35~UT to 04:55~UT.
Panels~(b) and (c) show the corresponding power spectra and the
global power spectra given by wavelet analysis, respectively.
Panels~(d) is the wavelet results obtained for the data averaged
(also normalized with trend subtraction) in C2 (see Figure~3) from
04:35~UT to 04:55~UT. The trend profile is obtained by taking
average of the original $T_B$ data every 400s. Subtracting the long-term average trend does not affect the
wavelet analysis result as long as the term is much longer than 2
mins. The horizontal dashed lines in the global spectra indicate
the significance level of 99{\%}. Cross-hatched regions on the
spectra show the cone of influence. Note that around 0.5-1~min, there exist some spectral power. However, the corresponding spectral features last only for few mins with much less power than the 2-min one, and the associated period varies in a relatively-large range from 0.5 to 1~min. We therefore suggest that these periodicities are not significant.\protect\\(A color version of
this figure is available in the online journal.)} \label{Fig4}
\end{figure}

\begin{figure}
\includegraphics[width=0.85\textwidth]{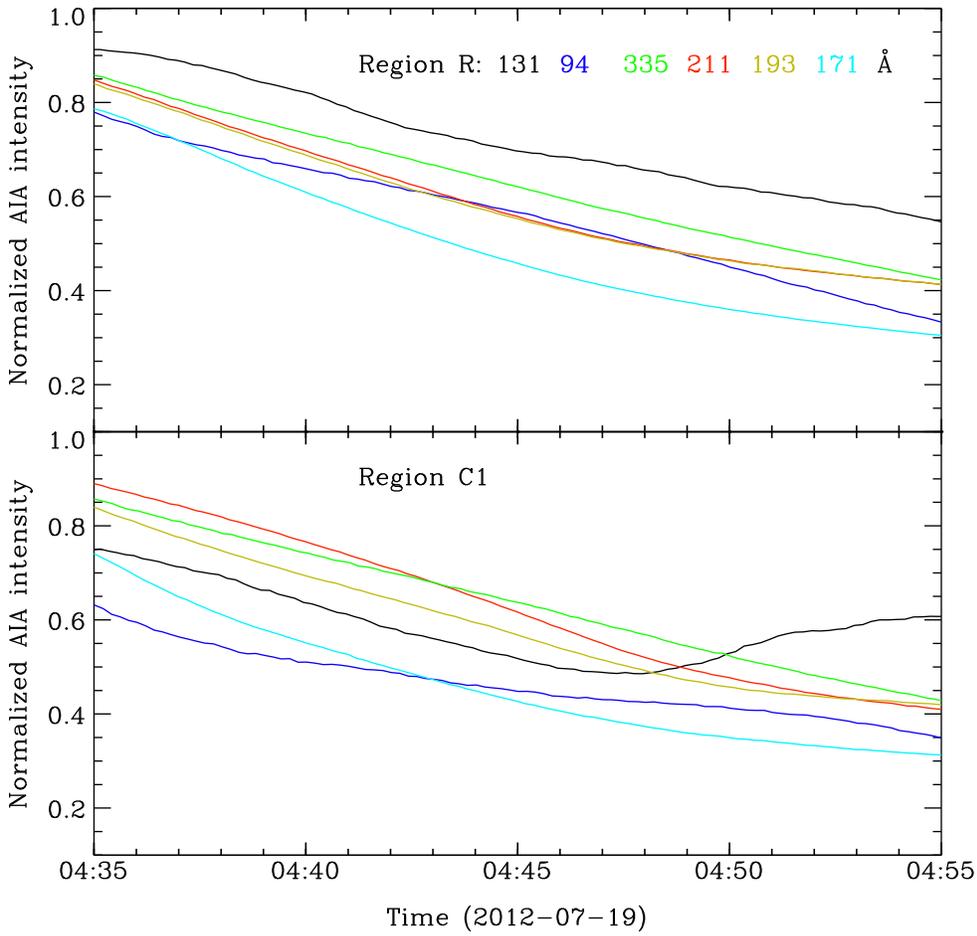}
\caption{AIA light curves at 131, 94, 335, 211, 193, and 171~\AA\
from 04:35 to 04:55~UT. The data are normalized to corresponding
values at 04:30~UT. \protect\\(A color version of this figure is
available in the online journal.)} \label{Fig5}
\end{figure}


\begin{thebibliography}{}
\bibitem[Aschwanden et al.(2013)]{Asch13}Aschwanden, M. J., Boerner, P., Schrijver, C. J., et al., 2013, \solphys, 283, 5
\bibitem[Battaglia \& Kontar(2012)]{Batta12}Battaglia, M., \& Kontar, E. P., 2012, \apj, 760, 142
\bibitem[Chen(1996)]{Chen96}Chen, J., 1996, J., Geophys. Res., 101, 27499
\bibitem[Cheng et al.(2013)]{Cheng13}Cheng, X., Zhang, J., Ding, M. D., 2013, \apj, 763, 43
\bibitem[Cheng et al.(2012)]{Cheng12}Cheng, X., Zhang, J., Saar, S.H., et al., 2012, \apj, 761, 62
\bibitem[Dulk(1985)]{Dulk85}Dulk, G. A., 1985, Ann. Rev. Astron. Astrophys., 23, 169
\bibitem[Forbes(2000)]{Forbes00}Forbes, T. G., 2000, J., Geophys. Res., 105, 23153
\bibitem[Kislyakova et al.(2011)]{Kis11}Kislyakova, K. S., Zaitsev, V. V., Urpo, S., et al., 2011, Astron. Rep., 55, 275
\bibitem[Krucker \& Battaglia(2014)]{Krucker14}Krucker, S., \& Battaglia, M., 2014, \apj, 780, 107
\bibitem[Kundu et al.(2001)]{Kundu01}Kundu, M. R., Nindos, A., \& White, S. M., 2001, \apj, 557, 880
\bibitem[Kundu et al.(2004)]{Kundu04}Kundu, M. R., Garaimov, V. I., White, S. M., et al., 2004, \apj, 600, 1052
\bibitem[Lemen et al.(2012)]{Lemen12}Lemen, J. R., Title, A. M., Akin, D. J., et al., 2012, \solphys, 275, 17
\bibitem[Lin et al.(2002)]{Lin02}Lin, R. P., Dennis, B. R., Hurford, G. J., et al., 2002, \solphys, 210, 3
\bibitem[Liu(2013)]{Liu13}Liu, R., 2013, \mnras, 434, 1309
\bibitem[Liu et al.(2013)]{Liu2013}Liu W., Chen Q., \& Petrosian V., 2013, \apj, 767, 168
\bibitem[Liu et al.(2012)]{Liu12}Liu, W., Ofman, L., Nitta, N. V., et al., 2012, \apj, 753, 52
\bibitem[Low(2001)]{Low01}Low, B. C., 2001, J., Geophys. Res., 106, 25141
\bibitem[Low \& Hundhausen(1995)]{Low96}Low, B. C., Hundhausen, J. R., 1995, \apj, 443, 818
\bibitem[Masuda et al.(1994)]{Masuda94}Masuda, S., Kosugi, T., Hara, H., et al., 1994, Nature, 371, 495
\bibitem[Morgachev et al.(2014)]{Mor14}Morgachev, A. S., Kuznetsov, S. A., \& Melnikov, V. F., 2014, Geomagnetism and Aeronomy, 54, 933
\bibitem[Nakajima et al.(1994)]{Nakajima94}Nakajima, H., Nishio, M., Enome, S., et al., 1994, IEEEP, 82, 705
\bibitem[Narukage et al.(2014)]{Narukage14}Narukage, N., Shimojo, M., \& Sakao, T., 2014, \apj, 787, 125
\bibitem[Ofman et al.(2011)]{Ofman11}Ofman, L., Liu, W., Title, A., et al., 2011, \apjl, 740, L33
\bibitem[Patsourakos et al.(2013)]{Patsourakos13}Patsourakos, S., Vourlidas, A., \& Stenborg, G., 2013, \apj, 764, 125
\bibitem[Pesnell et al.(2012)]{Pesnell12}Pesnell, W. D., Thompson, B. J., Chamberlin, P. C., et al., 2012, \solphys, 275: 3-15
\bibitem[Reznikova \& Shibasaki(2011)]{Reznikova11}Reznikova, V. E., \& Shibasaki, K., 2011, \aap, 525, 112
\bibitem[Song et al.(2015)]{Song15}Song, H. Q., Chen, Y., Zhang, J., et al., 2015, \apjl, 808, L15
\bibitem[Song et al.(2014)]{Song14}Song, H. Q., Zhang, J., Chen, Y., et al., 2014, \apj, 792, L40
\bibitem[Sun et al.(2014)]{Sun14}Sun J. Q., Cheng, X., Ding, M. D., 2014, \apj, 786, 73
\bibitem[Sych et al.(2009)]{Sych09}Sych, R., Nakariakov, V. M., Karlicky, M., et al., 2009, \aap, 505, 791
\bibitem[Takano et al.(1996)]{Takano96}Takano, T., Nakajima, H., Enome, S., et al, 1996, Proc. CESRA Workshop, eds. Trottet, G. Coronal physics from radio and space observations, Springer
\bibitem[Takano et al.(1997)]{Takano97}Takano, T., Nakajima, H., Enome, S., et al. 1997, LNP, 483, 183
\bibitem[White et al.(2011)]{White11}White, S. M., Benz, A. Q., Christe, S., et al., 2011, Space Sci. Rev., 159, 225
\bibitem[Yang et al.(2015)]{Yang15}Yang, L., Zhang, L., He, J., et
al., 2015, \apj, 800, 111
\bibitem[Zhang et al.(2012)]{Zhang12}Zhang, J., Cheng, X., Ding, M. D., 2012, NatCo, 3, 747
\end{thebibliography}
\end{document}